\documentclass[preprint,aps,showpacs,eqsecnum,nofootinbib]{revtex4}
\usepackage{amsmath}
\usepackage{amssymb}
\usepackage[dvips]{graphicx}
\usepackage{color}

\newcommand{\omits}[1]{}
\newcommand{\vect}[1]{\mbox{\boldmath $#1$}}

\newcommand{\diag}{\mathrm{diag}}
\newcommand{\inv}{^{-1}}
\newcommand{\half}{\frac 1 2}
\newcommand{\Mink}{\mathbb{R}^{1,4}}
\newcommand{\hypb}{\mathcal{S}_R}
\newcommand{\hypp}{\mathcal{P}}
\newcommand{\NH}{\mathcal{NH}}

\def\nno{\nonumber}

\def\Ga{\Gamma}

\def\al{\alpha}

\def\si{\sigma}

\def\bc{\begin{center}}
\def\ec{\end{center}}
\def\be{\begin{eqnarray}}
\def\ee{\end{eqnarray}}
\def\nno{\nonumber}

\def\del{\nabla}

\def\r{\partial}

\definecolor{dyellow}{rgb}{1.,0.8,.0}
\definecolor{myblue}{rgb}{.1,.1,.7}
\definecolor{dcyan}{rgb}{.0,.6,.6}
\definecolor{dmagenta}{rgb}{0.6,0.0,0.6}
\definecolor{brown}{rgb}{0.6,0.2,0.}
\definecolor{darkblue}{rgb}{.0,.0,0.5}
\definecolor{darkred}{rgb}{0.75,0.0,0.0}
\definecolor{orange}{rgb}{1.,.6,.0}
\definecolor{dorange}{rgb}{0.8,.4,.0}
\definecolor{darkgreen}{rgb}{0.0,0.6,0.0}
\definecolor{purple}{rgb}{.4,.0,.4}
\definecolor{lightgray}{rgb}{.8,.8,.8}

\begin{document}

\preprint{}

\title{Mechanics and Newton-Cartan-Like Gravity on the Newton-Hooke Space-time}

\author{{Yu Tian}$^{1}$} \email{ytian@itp.ac.cn}
\author{{Han-Ying Guo}$^{2,1}$} \email{hyguo@itp.ac.cn}
\author{{Chao-Guang Huang}$^{3}$} \email{huangcg@mail.ihep.ac.cn}
\author{{Zhan Xu}$^{4}$} \email{zx-dmp@mail.tsinghua.edu.cn}
\author{{Bin Zhou}$^{5,6}$} \email{zhoub@itp.ac.cn}

\affiliation{${}^1$ Institute of Theoretical Physics, Chinese
Academy of Sciences, P.O. Box 2735, Beijing 100080, China}

\affiliation{${}^2$ CCAST (World Laboratory), P.O. Box 8730,
Beijing 100080, China}

\affiliation{${}^3$ Institute of High Energy Physics, Chinese
Academy of Sciences, P.O. Box 918-4, Beijing 100049, China}

\affiliation{${}^4$ Physics Department, Tsinghua University,
Beijing 100084, China}

\affiliation{${}^5$ Physics Department, Beijing Normal University,
Beijing 100875, China}

\affiliation{$^6$ Interdisciplinary Center of Theoretical Studies,
Chinese Academy of Sciences, Beijing 100080, China}

\date{\today}

\begin{abstract}
We focus on the dynamical aspects on Newton-Hooke space-time
${\cal NH}_+$ mainly from the viewpoint of geometric contraction
of the de Sitter spacetime with Beltrami metric. (The term
spacetime is used to denote a space with non-degenerate metric,
while the term space-time is used to denote a space with
degenerate metric.) We first discuss the Newton-Hooke classical
mechanics, especially the continuous medium mechanics, in this
framework. Then, we establish a consistent theory of gravity on
the Newton-Hooke space-time as a kind of Newton-Cartan-like
theory, parallel to the Newton's gravity in the Galilei
space-time. Finally, we give the Newton-Hooke invariant
Schr\"odinger equation from the geometric contraction, where we
can relate the conservative probability in some sense to the mass
density in the Newton-Hooke continuous medium mechanics. Similar
consideration may apply to the Newton-Hooke space-time ${\cal
NH}_-$ contracted from anti-de Sitter spacetime.
\end{abstract}

\pacs{04.20.Cv, 45.20.-d, 02.40.Dr}

\maketitle

\tableofcontents

\newpage


\section{Introduction}

From the viewpoint of purely theoretical and fundamental physics,
it is well known that there are eight types of possible
kinematical symmetry groups based on some rather natural
assumptions \cite{Bacry}. Among them the most basic two are de
Sitter (dS) and anti-de Sitter (AdS) groups, which are $SO(1,d+1)$
and $SO(2,d)$ for $(1+d)$-dimensional spacetime, respectively; all
the others are In\"on\"u-Wigner contractions \cite{IW} of them.
The so-called Newton-Hooke (NH) group $N_\pm$ is an important and
interesting contraction of dS/AdS group, respectively. It is the
meaningful non-relativistic limit of dS/AdS group. At the same
time, the Galilei group is a further contraction (the flat limit)
of both $N_\pm$ groups. All these kinematical groups can lead to
the corresponding $(1+d)$-dimensional space-times as some
homogeneous spaces of them. Furthermore, the action of these
groups on their corresponding space-times can take some nice
fractional linear forms under special coordinate systems (called
Beltrami coordinates\footnote{Cartesian coordinates on the
(pseudo-)Euclidean spaces, actually, are the limiting case under
contraction of Beltrami coordinates.}), and the corresponding
mechanics, like Newtonian mechanics on Galilei space-time, can be
really established from first principles \cite{GHTXZ,GHTXZ4}.
Especially, the NH case as a non-relativistic cosmological
kinematics is studied in detail in \cite{DD} and recently in
\cite{GHTXZ}.

On the other hand, from the viewpoint of modern physics and
cosmology, constant curvature space-times have drawn much
attention, from both theoretical and observational considerations.
The significance of AdS space is early recognized, based upon the
fact that its symmetry algebra has supersymmetric extensions and
so it can be incorporated into supergravity and string theory.
Related study has resulted in the profound AdS/CFT correspondence
\cite{AdS/CFT}. The great interest in dS space comes from recent
cosmological observations showing that our universe is asymptotic
dS, \emph{i.e.}, with a positive cosmological constant
\cite{90s,WMAP}. However, there will be lots of puzzles within the
present framework of physics if our universe does have a positive
cosmological constant \cite{puzzle}. Under this embarrassed
situation, of course, any instructive attempts related to these
problems are worthwhile. One available attempt is just to consider
the non-relativistic limit, \emph{i.e.}, the NH limit, which
drastically simplifies the analysis while still taking the effects
of cosmological constant into account. That is why the interest in
NH space-time revives recent years \cite{ABCP,Gao,GP,GHTXZ}.

Following our recent paper \cite{GHTXZ} that investigates NH
space-time from the geometric contraction, in this paper we focus
on the dynamical aspects on NH space-time. We discuss in detail
the NH kinematics, dynamics and even continuous medium mechanics.
Especially, we establish a consistent theory of gravity on NH
space-time as a kind of Newton-Cartan-like theory, parallel to the
Newton's gravity on the Galilei space-time. We also discuss some
interesting aspects of the NH invariant Schr\"odinger equation. We
find that it is possible to relate the conservative probability in
some sense to the mass density in the NH continuous medium
mechanics. Unlike most of preceding articles, which investigate NH
mechanics mainly from the algebraic point of view, our discussion
will be more geometric and based on the foundation of physics. The
invariance of physics under the action of NH group plays an
important role in our discussion.

The paper is organized as follows. In Sec.\ref{sec:NH}, after a
brief introduction to algebraic construction of NH space-time, we
introduce the geometric description of dS/AdS spacetime, Beltrami
coordinates, dS/AdS group action and their NH limit for the dS
spacetime. In Sec.\ref{sec:mech} we discuss the NH mechanics,
concentrating on the dynamics. The Newton-Cartan-like theory on NH
space-time is constructed in Sec.\ref{sec:NC}. We give the
gravitational field equation there and solve it to obtain the law
of gravity for the exterior of spherical source. In
Sec.\ref{sec:Sch} we deduce the Schr\"odinger equation on NH
space-time from the geometric contraction, show its NH invariance,
and discuss the conservation of probability, which can be related
in some sense to the mass density in fluid mechanics. We end the
paper with a brief conclusion and discussion in Sec.\ref{sec:CD}.


\section{Newton-Hooke Space-time as a Limit of Beltrami-de Sitter Spacetime}
\label{sec:NH}

The Lie algebra of dS/AdS group, in terms of the time-space
decomposition, is (taking $d=3$ for definiteness) \cite{GHTXZ}
\begin{eqnarray}
&[{\bf J}_i,{\bf H}] = 0, \quad [{\bf J}_i,{\bf J}_j] =
\epsilon_{ijk}\mathbf{J}_k, \quad [{\bf J}_i,{\bf P}_j] =
\epsilon_{ijk}\mathbf{P}_k,& \nno \\
&[{\bf J}_i,{\bf K}_j] = \epsilon_{ijk}\mathbf{K}_k, \quad [{\bf
H},{\bf P}_i] = \pm\nu^2\mathbf{K}_i, \quad
[{\bf H},{\bf K}_i] = \mathbf{P}_i,& \label{dS} \\
&[{\bf P}_i,{\bf P}_j] = \pm R^{-2}\epsilon_{ijk}\mathbf{J}_k,
\quad [{\bf K}_i,{\bf K}_j] = -c^{-2}\epsilon_{ijk}\mathbf{J}_k,
\quad [{\bf P}_i,{\bf K}_j] = c^{-2}\delta_{ij}\mathbf{H},& \nno
\end{eqnarray}
where the generators have their usual meanings, $\nu:=c/R$ has the
same dimension as frequency, and the ``$+$"/``$-$" sign is for
dS/AdS, respectively. The Newton-Hooke (NH) algebra
$\mathfrak{n}_\pm (1,3)$ is the following limit (contraction) of
the above dS/AdS algebra:
\begin{eqnarray}\label{nu}
c \to \infty, \quad R\to \infty, \quad {\rm but}\quad
  \nu = \frac c R {\rm ~ is ~ a ~ positive, ~ finite ~ constant},
\end{eqnarray}
which reads
\begin{eqnarray}
&[{\bf J}_i,{\bf J}_j] = \epsilon_{ijk}\mathbf{J}_k, \quad [{\bf
J}_i,{\bf P}_j] =
\epsilon_{ijk}\mathbf{P}_k,& \nno \\
&[{\bf J}_i,{\bf K}_j] = \epsilon_{ijk}\mathbf{K}_k, \quad [{\bf
H},{\bf P}_i] = \pm\nu^2\mathbf{K}_i, \quad
[{\bf H},{\bf K}_i] = \mathbf{P}_i,& \label{NH algebra} \\
&{\rm and ~ the ~ other ~ Lie ~ brackets ~ vanish.}& \nno
\end{eqnarray}
If we first replace $\mathbf{H}$ with $\mathbf{H}-\mathrm{i}m
c^2$, where $m$ is a central element, and then perform the
contraction, we will get the central extension
$\mathfrak{n}_\pm^\mathrm{C}(1,3)$ of NH algebra (\ref{NH
algebra}):
\begin{eqnarray}
&[{\bf J}_i,{\bf J}_j] = \epsilon_{ijk}\mathbf{J}_k, \quad [{\bf
J}_i,{\bf P}_j] =
\epsilon_{ijk}\mathbf{P}_k,& \nno \\
&[{\bf J}_i,{\bf K}_j] = \epsilon_{ijk}\mathbf{K}_k, \quad [{\bf
H},{\bf P}_i] = \pm\nu^2\mathbf{K}_i, \quad
[{\bf H},{\bf K}_i] = \mathbf{P}_i,& \label{NH algebra ext} \\
&[{\bf P}_i,{\bf K}_j] = -\mathrm{i}\delta_{ij}m, \quad {\rm and ~
the ~ other ~ Lie ~ brackets ~ vanish.}& \nno
\end{eqnarray}
From the spacetime point of view, the parameter $R$ in
eq.(\ref{dS}) is the cosmic radius, which is related to the
cosmological constant $\Lambda$ by $\Lambda=\pm 3R^{-2}$. Now in
the NH limit the new parameter $\nu$ takes its place and has the
meaning of temporal curvature. If we perform a further contraction
$\nu\to 0$ (the so-called flat limit) the algebras
$\mathfrak{n}_\pm$ and $\mathfrak{n}_\pm^\mathrm{C}$ come back to
the familiar Galilei algebra $\mathfrak{gal}$ and the
corresponding central extension $\mathfrak{gal}^\mathrm{C}$,
respectively. The algebra $\mathfrak{gal}^\mathrm{C}$ is well
known as the symmetry of non-relativistic quantum mechanics
(Schr\"odinger equation). And it can be seen there that the
central element $m$ corresponds to the mass.

The above statements on the NH limit is from the Lie algebra point
of view (or after exponentiating, from the Lie group point of
view). But the group aspects are far from sufficiency. The
geometric aspects, such as connections, metrics (if exist) etc,
are very important when concerning physics on these space-times.
Conventionally, the next step is to consider dS and AdS spacetimes
and NH space-time as homogeneous spaces $SO(1,d+1)/SO(1,d)$,
$SO(2,d)/SO(1,d)$ and $N_\pm (1,d)/\hat{N}_\pm (1,d)$,
respectively, while considering the original groups as the
corresponding principal bundles over them. Here $\hat{N}_\pm
(1,d)$ is the homogeneous NH group, whose Lie algebra is the
subalgebra of $\mathfrak{n}_\pm (1,d)$ generated by $\mathbf{J}_i$
and $\mathbf{K}_i$.\footnote{In fact, it is easy to see that
$\hat{N}_\pm (1,d)$ and the homogeneous Galilei group are both
isomorphic to $SO(d)\otimes_\mathrm{S}\mathbb{R}^d$.} Then one can
examine the actions of dS, AdS and NH groups on these homogeneous
spaces. In this picture the invariant connections on these spaces
can be systematically obtained as the so-called ``canonical"
connections \cite{KN,ABCP}. However, this picture does not help us
establish the NH dynamics when taking into account the gravity
from matter.

Fortunately, the dS/AdS spacetime has a simple geometric
description as the pseudo-sphere embedded in higher dimensional
Minkowski spacetime. The NH space-time
$\mathcal{NH}_\pm$ can be directly obtained as
some appropriate limit of this geometric picture. In fact, this
naive limiting procedure can give us all the necessary geometric
information of NH space-time. So, from now on, we can forget the
algebraic construction of NH space-time and study
$\mathcal{NH}_\pm$ directly from a geometric point of view. In the
following, we only consider the dS case and the corresponding
$\mathcal{NH}_+$ (denoted by $\mathcal{NH}$ for briefness).
$\mathcal{NH}_-$ can be dealt with in parallel.


\subsection{Hyperboloid Model of de Sitter Spacetime}

\begin{figure}[!hbt]
\begin{center}
  \includegraphics[width=75mm,height=75mm]{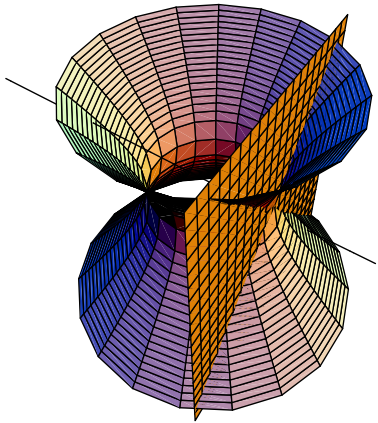}
  \caption{The hyperboloid $\hypb$ and the hyperplane $\hypp$.
  The Straight line passes through 0 and a pair of antipodal points in $\hypb$.}
  \label{fig:dS}
\end{center}
\end{figure}

As is well known, the 4-dimensional dS spacetime can be viewed as
a hyperboloid (Fig.\ref{fig:dS})
\begin{equation}\label{eq:hyperboloid}
  \hypb: \eta_{AB}\xi^A\xi^B = - R^2, \quad
  \eta_{AB}=\diag(1,-1,-1,-1,-1),
\end{equation}
with topology $S^3\times\mathbb{R}$, in the 5-dimensional
Minkowski spacetime $\Mink$. Indices $A$, $B$, etc., run over 0 to
4, while Greek indices such as $\mu$, $\nu$ run over 0 to 3.

Since $\hypb$ is invariant under the action of $O(1,4)$ on
$\Mink$, the latter induces an action on $\hypb$. This
transformation group is called the dS group. In this paper
we are mainly interested in the invariant structure of dS
spacetime under the action of $SO^\uparrow(1,4)$, the connected
Lie subgroup of $O(1,4)$ that preserves the orientation and time
orientation. It is denoted by $G$.

The next step is to chose some coordinate systems on the dS
spacetime, which remain meaningful after the NH limit, and which
admit a physical interpretation (as inertial frames, actually) and
can be used to establish kinematics on dS spacetime and NH
space-time. For these reasons, and in order to relate our
discussion to physical principles \cite{GHTXZ} in future works, we
choose the Beltrami coordinates.

\subsection{Beltrami-de Sitter Model}
\label{sec:BdS}

\subsubsection{Beltrami Coordinates}

Now let $\hypp_4^+$ be the hyperplane $\xi^4 = R$ in $\Mink$. For
each point $\xi\in\hypb$ with $\xi^4>0$, there is a
one-to-one-corresponding point $x\in\hypp_4^+$ such that
\begin{equation}\label{Beltrami}
\xi_x^\mu = R\frac{\xi^\mu}{\xi^4}=:x^\mu, \quad \xi_x^4 = R.
\end{equation}
This map is actually obtained by drawing a straight line passing
through $\xi$ and $0\in\Mink$, as shown in Fig.\ref{fig:dS}. Since
$\xi\in\hypb$ satisfies eq.(\ref{eq:hyperboloid}), the
corresponding $x$ in $\hypp_4^+$ satisfies
\begin{equation}\label{eq:ineql}
  \sigma(x) > 0,
\end{equation}
where
\begin{equation}\label{eq:sigma}
  \sigma(x) = 1 - R^{-2}\eta_{\mu\nu}x^\mu x^\nu.
\end{equation}
The above ``gnomonic" projection from $\hypb$ with $\xi^4>0$ into
$\hypp_4^+$ defines a coordinate system on a patch, denoted
$U_4^+$, of $\hypb$, which is known as a Beltrami coordinate
system \cite{GHXZ,GHXZ2,GHTXZ}. Note that in order to preserve the
orientation, the antipodal identification should not been taken.
Under the Beltrami coordinates, the metric on $\hypb$ has the form
\begin{equation}\label{proper}
ds^2=[\eta_{\mu\nu}\sigma^{-1}(x)+R^{-2}\eta_{\mu\rho}\eta_{\nu\sigma}x^\rho
x^\sigma \sigma^{-2}(x)]dx^\mu dx^\nu.
\end{equation}
We also give the Christoffel connection
\begin{equation}\label{Christoffel}
\Gamma^{\phantom{\mu}\rho}_{\mu\nu}
  =  \frac{x_\mu\delta^\rho_\nu + x_\nu\delta^\rho_\mu}{R^2\sigma(x)}
\end{equation}
of this metric for later reference.

The patch $U_4^+$ covers almost half of $\hypb$. The other half is
almost covered by another patch $U_4^-$, which is the ``gnomonic"
projection from $\hypb$ with $\xi^4<0$ into the hyperplane
$\hypp_4^-$ located at $\xi^4 = -R$ in $\Mink$. The Beltrami
coordinates on $U_4^-$ is given by
\begin{equation}
  x^\mu = -R\frac{\xi^\mu}{\xi^4}.
\end{equation}
Obviously, there are at least eight patches
$U_\alpha^\pm,\alpha=1,\cdots,4$ to cover the whole $\hypb$. In
patches $U_\alpha^\pm,\alpha=1,2,3$, the Beltrami coordinates are
given by
\begin{equation}
  x^\nu = \pm R\frac{\xi^\nu}{\xi^\alpha},\quad
  \nu=0,\cdots,\hat{\alpha},\cdots,4,\quad \xi^\alpha\gtrless 0,
\end{equation}
where $\hat{\alpha}$ means omission of $\alpha$. In the following
discussions, we mainly concentrate on the $U_4^+$ patch.

The 3-dimensional hyperboloid
\begin{equation}\label{eq:scri}
\sigma(x)=0
\end{equation}
is a part of the projective boundary of $\hypb$ \cite{GHTXZ4},
which corresponds to the conformal boundary on the Penrose diagram
of dS spacetime. In fact, it is the intersection of $\hypp_4^+$
and the 5-dimensional light cone
\begin{equation}
  \eta_{AB}\xi^A\xi^B = 0.
\end{equation}

\subsubsection{Fractional Linear Form of de Sitter Group}

The isometry group of $\hypb$ is $O(1,4)$. Its subgroup
$SO^\uparrow(1,4)$ which preserves the orientation and time
orientation of $\hypb$ has been denoted by $G$. Let $(D^A_{\
B})\in G$. Then a point $(\xi^A)\in\Mink$ will be sent to another
point $(\xi'^A) = (D^A_{\ B}\,\xi^B)$. Examples show that $D^4_{\
4}$ can be arbitrary real number when $(D^A_{\ B})$ runs over $G$.
Later we will identify the transformations in $G$ with
transformations among inertial frames on dS spacetime (and its NH
limit $\mathcal{NH}$).

For a given $(D^A_{\ B})$, if $D^4_{\ 4} \neq 0$,\footnote{The
$D^4_{\ 4} = 0$ case is a little subtle, which is discussed in
\cite{GHTXZ4}.} we can define
\begin{equation}
  a_\mu = - R\,\frac{D^4_{\ \mu}}{D^4_{\ 4}},\qquad
  a^\mu = \eta^{\mu\nu} a_\nu.
\end{equation}
Then we can obtain the relations
\begin{equation}
  D^4_{\ 4} = \pm \sigma^{-1/2}(a),\qquad
  D^\mu_{\ 4} = - \frac{D^\mu_{\ \nu}\,a^\nu}{R}.
\end{equation}
In the following, the signs $\pm$ and $\mp$ are taken
corresponding to the sign of $D^4_{\ 4}$. We can define
\begin{equation}
  L^\mu_{\ \nu} = D^\mu_{\ \nu}
  - \frac{R^{-2}\,D^\mu_{\ \rho} \, a^\rho a_\nu}{ 1 + \sigma^{1/2}(a)},
\end{equation}
which satisfy
\begin{equation}\label{ortho}
\eta_{\rho\sigma}L^\rho_{\ \mu} L^\sigma_{\ \nu} = \eta_{\mu\nu}.
\end{equation}
The inverse relation reads
\begin{equation}\label{D}
  D^\mu_{\ \nu} = L^\mu_{\ \nu}
  + \frac{R^{-2}\,L^\mu_{\ \rho}\,a^\rho a_\nu}{\sigma(a) + \sigma^{1/2}(a)}.
\end{equation}

The action of $(D^A_{\ B})$ on $U_4^+$ can be easily obtained from
eq.(\ref{Beltrami}). The result takes a factional linear form:
\begin{equation}\label{eq:D2D}
  \tilde{x}^\mu
  = \pm \frac{B^\mu_{\ \nu} (x^\nu - a^\nu)}{\sigma(a,x)},
\end{equation}
where
\begin{equation}
  \sigma(a,x): = 1 - R^{-2}\,\eta_{\mu\nu} a^\mu x^\nu,\qquad
  B^\mu_{\ \nu}:=\sigma^{1/2}(a)D^\mu_{\ \nu}.
\end{equation}
In fact, if $\pm\sigma(a,x) > 0$, $x'$ remains in the $U_4^+$
patch and so eq.(\ref{eq:D2D}) is valid; if $\pm \sigma(a,x) \leq
0$, then $x'$ will go out of $U_4^+$ and a transition between
coordinate patches is needed. It is important that all transition
functions in intersections can be realized by elements of $G$,
which is easily understood.


\subsection{The Newton-Hooke Limit}
\label{sec:limit}

Here we choose the most convenient way to consider the NH limit
from the 5-dimensional point of view. Replace for all equations in
Sec.\ref{sec:BdS} the original metric $\eta_{AB}$ with
\begin{equation}
g_{AB}=\diag(c^2,-1,-1,-1,-1),
\end{equation}
where $c$ is a positive constant with the physical meaning speed
of light. Now $\xi^0$ has a dimension of time, and we can define
\begin{equation}
t\equiv x^0=R\xi^0/\xi^4.
\end{equation}
When $c$ increases, the 5-dimensional light cone
\begin{equation}
g_{AB}\xi^A\xi^B=0
\end{equation}
collapses. The NH limit is attained when $c,R\rightarrow\infty$
while keeping $\nu\equiv c/R$ fixed. This will keep the crossing
points of the 5-d light cone and the $x^0$ axis on $\hypp_4^+$
fixed at $x^0=\pm 1/\nu$. In fact, the $c\rightarrow\infty$ limit
of the original 5-d Minkowski spacetime is the 5-d Galilei
space-time. The latter has a degenerate (or split) space-time
metric, which induces the split metric of NH space-time
$\mathcal{NH}$. Now the $U_4^+$ patch is itself geodesically
complete, so other coordinate patches are no longer needed. It is
easy to see that the projective boundary becomes the hyperplanes
$t=\pm 1/\nu$ in $\mathcal{NH}$.

Now put the NH limit in a little more detail. Using
$g_{\mu\nu}=\diag(c^2,-1,-1,-1)$, eq.(\ref{eq:sigma}) becomes
\begin{equation}
\sigma(t)=1-\nu^2 t^2
\end{equation}
under the NH limit. From the metric $g_{\mu\nu}$ the Lorentz
matrix $(L^\mu_{\ \nu})$ has the familiar Newtonian limit:
\begin{equation}\label{Newton}
\left(%
\begin{array}{cc}
  L^0_{\ 0} & L^0_{\ j} \\
  L^i_{\ 0} & L^i_{\ j} \\
\end{array}%
\right)\rightarrow\left(%
\begin{array}{cc}
  1 & 0 \\
  -O^i_{\ j}u^j & O^i_{\ j} \\
\end{array}%
\right), \quad O^i_{~j} \in SO(3).
\end{equation}
Correspondingly, one can obtain the NH limit of $D^\mu_{\ \nu}$
from eq.(\ref{D}):
\begin{eqnarray}
D^0_{\ 0} & \rightarrow & \frac{1}{\sigma^{1/2}(t_a)}, \\
D^0_{\ j} & \rightarrow & 0, \\
D^i_{\ 0} & \rightarrow & - \frac{O^i_{\ j}u^j}{\sigma^{1/2}(t_a)}
  + \frac{\nu^2 t_a O^i_{\ j}a^j}{\sigma(t_a) +
  \sigma^{1/2}(t_a)} =: -O^i_{\ j}\hat{u}^j, \\
D^i_{\ j} & \rightarrow & O^i_{\ j},
\end{eqnarray}
where $t_a\equiv a^0$. Hereafter, $\hat{\vect{u}}$ is renamed to
$\vect{u}$ for convenience.

Because the restriction to $U_4^+$ requires $D^4_{\ 4}>0$, we will
have from eq.(\ref{eq:D2D})
\begin{eqnarray}
\tilde{t}&=&\frac{t - t_a}{\sigma(t_a,t)}, \\
\tilde{x}^i&=&\frac{\sigma^{1/2}(t_a)}{\sigma(t_a,t)}O^i_{\ j}[x^j
- a^j - u^j (t - t_a)].
\end{eqnarray}
Defining
\begin{equation}
b^i\equiv a^i - u^i t_a,
\end{equation}
the above transformation becomes the same form as in \cite{GHTXZ}:
\begin{eqnarray}
\tilde{t}&=&\frac{t - t_a}{\sigma(t_a,t)}, \label{NH t} \\
\tilde{x}^i&=&\frac{\sigma^{1/2}(t_a)}{\sigma(t_a,t)}O^i_{\ j}(x^j
- b^j - u^j t). \label{NH x}
\end{eqnarray}
The group properties of this type of fractional linear
transformations have been discussed in \cite{GHTXZ}. It is
important that the transformation for time coordinate is
independent of space coordinates, and that the transformation for
space coordinates are linear among themselves. Thus it follows
that the Beltrami-time simultaneity on $\mathcal{NH}$ is absolute,
\emph{i.e.}, independent of (inertial) reference frames, which is
similar to the Newtonian space-time.

Considering the infinitesimal form of NH transformation (\ref{NH
t},\ref{NH x}), we can get the Beltrami-coordinate realization of
(anti-Hermitian) generators of the NH algebra $\mathfrak{n}_+
(1,3)$:
\begin{eqnarray}
&{\bf H}=\sigma(t)\partial_t -\nu^2 t x^i\partial_i,& \nno \\
&{\bf P}_i=\partial_i, \quad {\bf K}_i = t\partial_i,& \label{realize} \\
&{\rm and ~ the ~ usual ~ form ~ of ~ the ~ } SO(3) {\rm ~
generators ~ } \mathbf{J}_i.& \nno
\end{eqnarray}
Then the Lie brackets (\ref{NH algebra}) are easily checked.

The meaningful NH limit of eq.(\ref{proper}) is
\begin{equation}\label{dtau}
d\tau^2 = c^{-2}ds^2 = \sigma^{-2}(t) dt^2.
\end{equation}
If $d\tau^2$ is taken as the new line element instead of $ds^2$,
we will have the following degenerate metric tensor:
\begin{equation}\label{metric}
g_{tt}=\sigma^{-2}(t), \qquad g_{ij}=0, \qquad g_{ti}=g_{it}=0.
\end{equation}
In a fixed hypersurface of simultaneity ($dt=0$), we have
\begin{equation}\label{l}
dl^2=\hat{g}_{ij}dx^i dx^j, \quad
\hat{g}_{ij}=\si^{-1}(t)\delta_{ij}.
\end{equation}
A connection exists as the contraction of the Christoffel
connection (\ref{Christoffel}), whose nonzero coefficients are
only
\begin{equation}\label{affine}
  \Gamma^{\ t}_{tt} = \frac{2\nu^2 t}{1 - \nu^2 t^2}, \qquad
  \Gamma^{\ i}_{tj} = \Gamma^{\ i}_{jt}
  = \frac{\nu^2 t}{1 - \nu^2 t^2}\delta^j_i.
\end{equation}
It is pleasant to see that this connection is torsion-free, as
expected, and that the corresponding curvature tensor and Ricci
tensor have the following nonzero components:
\begin{equation}\label{curv}
  R^i_{t\mu\nu} = \frac{\nu^2}{(1 - \nu^2 t^2)^2}\,
  (\delta^t_\mu \delta^i_\nu - \delta^i_\mu \delta^t_\nu)
\quad\textrm{and}\quad
  R_{tt} = \frac{-3\nu^2}{(1 - \nu^2 t^2)^2},
\end{equation}
respectively. Note that eq.(\ref{curv}) can be directly obtained
by contracting the curvature tensor and Ricci tensor on $\hypb$,
and that the relation
\begin{equation}\label{empty NC}
R_{tt} = -3\nu^2 g_{tt}
\end{equation}
holds as expected.

It is easy to check that eqs.(\ref{dtau}), (\ref{metric}),
(\ref{l}), (\ref{affine}) and (\ref{curv}) are all invariant under
NH transformations. It can be proved that the above connection is
the only one that is NH invariant and keeps $d\tau$ invariant. For
details, see Appendix \ref{app:NHinv}. Further, under the flat
limit $\nu\to 0$, all the above expressions reduce to their
counterparts in the Newtonian case.


\section{Newton-Hooke Classical Mechanics}
\label{sec:mech}


\subsection{Newton-Hooke Kinematics}

Following \cite{GHTXZ}, we only list here some related results of
the kinematics on NH space-time. Differentiating NH transformation
(\ref{NH t},\ref{NH x}) gives rise to the velocity composition law
\begin{equation}\label{v trans}
\tilde{v}^i=\frac{O_{\
j}^i}{\si^{1/2}(t_a)}[\si(t_a,t)v^j-u^{j}+\nu^{2}t_a
(x^{j}-b^{j})]
\end{equation}
for $\vect{v}\equiv d\vect{x}/dt$. Differentiating again, one
obtains the following transformation of (3-)acceleration:
\begin{equation}\label{a trans}
\frac{d\tilde{v}^i}{d\tilde{t}}=\frac{\sigma^3(t_a,t)}{\sigma^{3/2}(t_a)}O^i_{\
j}\frac{d v^j}{dt}.
\end{equation}
Surprisingly, the NH transformation of acceleration is much
simpler than that of velocity.

Noting that the NH transformation (\ref{v trans}) of velocity is
dependent on the position $\vect{x}$, we can define a new
quantity
\begin{equation}
V^i\equiv v^i+\frac{\nu^2 t x^i}{1-\nu^2 t^2},
\end{equation}
whose NH transformation is independent of $\vect{x}$:
\begin{equation}\label{Vtrans}
\tilde{V}^i=\frac{\si(t_a,t)}{\si^{1/2}(t_a)}O_{\ j}^i
\Big[V^j-\frac{u^j}{\si(t)}-\frac{\nu^2 t b^j}{\si(t)}\Big].
\end{equation}
We will see later that this quantity is very useful.


\subsection{Newton-Hooke Dynamics}

It is well-known that the gnomonic projection maps a great circle
(also a geodesic) on a sphere to a straight line on the target
plane. Since the dS/AdS spacetime is a pseudo-sphere (see
eq.(\ref{eq:hyperboloid}) for the dS spacetime), one can expect
that the similar conclusion holds. This is indeed the case, and is
actually an important reason why we chose such a kind of
coordinate systems \cite{GHTXZ5}. Based on this, we can define the
inertial motion (free motion, or moving along geodesics) as
uniform-velocity motion, parallel to the corresponding concept in
Newtonian mechanics and Special Relativity, and identify the
Beltrami coordinates with inertial frames. The fractional linear
transformation (\ref{eq:D2D}) preserves straight (world) lines.
Then the whole mechanics on dS/AdS spacetime can be established.
In fact, it is more appropriate to examine this from a
projective-geometry-like point of view \cite{GHTXZ4}, which we
will not dwell on in this article.

In the NH limit, one can intuitively expect from the geometric
picture that via Beltrami coordinates the relation between
geodesics and straight lines survives. This expectation can be
strictly proved using the geodesic equation with connection
(\ref{affine}) \cite{GHTXZ}. Thus, we have the counterpart of
Newton's first law on $\mathcal{NH}$, which we call Newton-Hooke's
first law.

To go further along this direction, we first list the (conserved)
non-relativistic energy and 3-momentum obtained in \cite{GHTXZ} as
\begin{eqnarray}
E_\mathrm{k} & = & \half m\vect{v}^2 - \frac{m\nu^2}{2}
(\vect{x}-t\vect{v})^2, \label{energy} \\
\vect{P} &=& m\vect{v}. \label{momentum}
\end{eqnarray}
Then, to justify that we can extend Newton's second law
\begin{equation}\label{Newton 2nd}
\frac {dP^i}{dt}=F^i
\end{equation}
to the NH space-time, it is expected that at least one side of the
above equation has good property under NH transformation (\ref{NH
t},\ref{NH x}). In fact, we see from eq.(\ref{momentum}) that the
transformation property of $dP^i/dt$ is the same as that of
acceleration (\ref{a trans}). So if we assume that the force $F^i$
has the same transformation property, Newton's second law can hold
on $\mathcal{NH}$, which we call Newton-Hooke's second law.

Differentiating the kinetic energy-momentum relation
\begin{equation}
E_\mathrm{k}=\frac 1 {2m}
\vect{P}^2-\frac{\nu^2}{2m}(m\vect{x}-t\vect{P})^2
\end{equation}
obtained from eqs.(\ref{energy},\ref{momentum}), we have
\begin{equation}\label{kinetic}
dE_{\mathrm{k}}=(1-\nu^{2}t^{2})\vect{F}\cdot
d\vect{x}+\nu^{2}t\vect{x}\cdot \vect{F}dt.
\end{equation}
This can be regarded as the kinetic energy theorem in $\NH$. A
detailed discussion on the kinetic and potential energy can be
found in Appendix \ref{app:E}.

Since the NH group $N_\pm (1,d)$, similar to the Galilei group,
has the space-translation subgroup $\mathbb{R}^d$, one can expect
that the conservation law of momentum (for a system of particles),
or equivalently Newton's third law, is respected in some sense. In
fact, it is easy to show from the velocity composition law (\ref{v
trans}) that for a two-body system the usual definitions
\begin{equation}
m=m_1+m_2, \qquad \vect{p}=\vect{p}_1+\vect{p}_2
\end{equation}
are invariant under NH transformations, which can be generalized
to many-body systems. The conservation of total momentum will lead
to the reversion of acting and reacting forces, which again we
call Newton-Hooke's third law. Later we will see that for the
gravitational interaction on $\mathcal{NH}$ Newton-Hooke's third
law is really respected.

\subsection{On Newton-Hooke Continuous Medium Mechanics }

In a general curved spacetime, we have the covariant conservation
of stress-energy tensor,
\begin{equation}\label{co cont}
D^\mu T_{\mu\nu}=g^{\mu\beta}(\partial_\beta T_{\mu\nu}-\Gamma^{\
\alpha}_{\mu\beta}T_{\alpha\nu}-\Ga_{\nu\beta}^{\ \al}T_{\mu
\al})=0.
\end{equation}
In the present paper, we use $D^\mu$ denoting the covariant
derivative and $\del$ the derivative operator in 3-space. Now
considering the dS spacetime, we substitute
eqs.(\ref{proper},\ref{Christoffel},\ref{eq:D2D}) into the above
equations. Under the NH limit, the temporal component of
eq.(\ref{co cont}) becomes the equation of continuity:
\begin{equation}\label{continuity}
\sigma(t)\partial_t\frac{\varrho}{\si^2(t)}-\nu^2 t
x^i\partial_i\frac{\varrho}{\si^2(t)}-4\nu^2
t\frac{\varrho}{\si^2(t)}+\partial_i\frac{\jmath^i}{\si(t)}=0,
\end{equation}
where we have defined $T_{tt}=\si^{-2}(t)\varrho$ and
$T_{it}=-c^{-2}\si\inv(t)\jmath^i$. Taking the NH limit of
coordinate transformation law of the stress-energy tensor,
\begin{equation}\label{T trans}
\tilde{T}_{\mu\nu}=\frac{\r
x^\alpha}{\r\tilde{x}^\mu}T_{\alpha\beta}\frac{\r
x^\beta}{\r\tilde{x}^\nu},
\end{equation}
we see that $\varrho$ is a scalar under NH transformations and
$\vect{\jmath}$ transforms as\footnote{The first order Newtonian
limit (\ref{Newton}) is not enough for considering the NH
transformation of $\vect{\jmath}$. One must carefully retain terms
of order $c^{-2}$ in $L^0_{\ j}$.}
\begin{equation}\label{j trans}
\tilde{\jmath}^i=\frac{\si(t_a,t)}{\si^{1/2}(t_a)}O_{\ j}^i
\Big[\jmath^j-\varrho\frac{u^j}{\si(t)}-\varrho\frac{\nu^2 t
b^j}{\si(t)}\Big],
\end{equation}
which is similar to eq.(\ref{Vtrans}). If we further define
\begin{equation}\label{hat rho}
\hat{\rho}=\si^{-3/2}(t)\varrho
\end{equation}
and
\begin{equation}\label{hat j}
\hat{\vect{j}}=\sigma^{-3/2}(t)\vect{\jmath}-\sigma^{-5/2}(t)\nu^2
t\vect{x}\varrho,
\end{equation}
eq.(\ref{continuity}) will become the same form
\begin{equation}\label{flat cont}
\partial_{t}\hat{\rho}+\nabla\cdot\hat{\vect{j}}=0
\end{equation}
as in the flat spaces.

The stress-energy tensor for a perfect fluid is
\begin{equation}
T_{\mu\nu}=(\varrho+p)U_\mu U_\nu-p g_{\mu\nu},
\end{equation}
where $U_\mu$ is the 4-velocity. The covariant conservation
(\ref{co cont}) gives rise to
\begin{eqnarray}
U_\mu D^\mu\varrho+(\varrho+p)D^\mu U_\mu &=& 0, \label{conteq0} \\
(\varrho+p)U_\mu D^\mu U_\nu+(U_\mu U_\nu-g_{\mu\nu})D^\mu p &=&
0. \label{euleq0}
\end{eqnarray}

Considering the dS spacetime and taking the NH limit, we obtain
from eq.(\ref{conteq0}) the same equation of continuity (\ref{flat
cont}) if $\hat{\rho}$ is still given by eq.(\ref{hat rho}) and
$\hat{\vect{j}}$ now given by
\begin{equation}
\hat{j}^{i}=\frac{\varrho v^i}{\si^{3/2}(t)}=\frac{\varrho
V^i}{\si^{3/2}(t)}-\frac{\nu^2 t x^i\varrho}{\si^{5/2}(t)},
\end{equation}
where $\vect{v}$ and $\vect{V}$ now stand for the velocity fields
of the NH perfect fluid. Comparing this expression with
eq.(\ref{hat j}), one sees that
\begin{equation}
\vect{\jmath}=\varrho\vect{V}
\end{equation}
for NH perfect fluid, which is consistent with the fact that both
sides of this equation have exactly the same NH transformation
property.

It can be shown that eq.(\ref{euleq0}) becomes
\begin{equation}
\varrho (\r_t v^i+v^j\r_j v^i)=-\si\inv(t)\r_i p
\end{equation}
or
\begin{equation}
\hat{\rho} (\r_t v^i+v^j\r_j v^i)=-\si^{-5/2}(t)\r_i p
\end{equation}
under the NH limit, which is the Euler equation for a perfect
fluid on $\NH$. It is also straightforward to check the NH
invariance of this equation.


\section{Newton-Cartan-Like Theory on the Newton-Hooke Space-time}
\label{sec:NC}

It is well known that Newton's gravity can be formulated in
torsion-free affine spaces \cite{Havas,Kunzle} due to Cartan's
observation \cite{Cartan}. In Sec. \ref{sec:limit}, $\mathcal{NH}$
has been shown to be a torsion-free affine space with nonzero
curvature. (See eq.(\ref{affine}) for the connection
coefficients.) So one may naturally expect that some kind of
Newton-Cartan theory can be constructed to describe the
gravitational interaction on $\mathcal{NH}$. In the present
section, we try to follow Cartan to set up a self-consistent
Newton-Cartan-like theory of gravitational interaction on $\NH$.
As a simple dynamical model that takes the effect of cosmological
constant into account, this theory may be valuable in the study of
cosmology.

To construct any self-consistent, dynamical theories on
$\mathcal{NH}$, the formulation of physical laws (or in other
words, dynamical equations) should be invariant under NH
transformations. Note that the connection in any
Newton-Cartan-like theory will not be NH invariant because in the
spirit of Newton-Cartan theory matter modifies the connection on
the space-time and because the NH invariant connection has been
determined up to a constant (See Appendix \ref{app:NHinv}).
Similar to the Newtonian case, it can be shown that the
Newton-Cartan-like connection cannot be fully determined by the
invariance of physical laws and the gravitational field equation.
Therefore, in order to obtain a unique and simple description of
the Newton-Cartan-like theory, what we shall do in the following
is to introduce physical requirements to preserve the NH
invariance of as many as possible coefficients of the connection.


\subsection{Physical Requirements and Gravitational Field
Equation}

Following the Newton-Cartan theory, we require that a test
particle in gravitational field moves along a geodesic with
respect to the Newton-Cartan-like connection. We also require,
from physical considerations, that the absolute time on empty
$\mathcal{NH}$ is preserved, and that Newton-Hooke's second law is
valid for gravitational action. For simplicity, the
Newton-Cartan-like connection coefficients are still denoted by
$\Gamma_{\mu\nu}^{\ \rho}$.

First, when the absolute time is not affected by the introduction
of interactions, including gravity, we have from eq.(\ref{dtau})
\begin{equation}
  \frac{d^{2}t}{d\tau^{2}}
  + \frac{2\,\nu^{2}t}{1-\nu^2 t^2}\frac{dt}{d\tau}\frac{dt}{d\tau}=0.
\end{equation}
Second, the Newton-Hooke's second law (\ref{Newton 2nd}) can be
rewritten as
\begin{equation}
  \frac{d^{2}x^{i}}{d\tau^{2}}
  = \frac{F^i}{m}\frac{dt}{d\tau}\frac{dt}{d\tau}
  - \frac{2\nu^{2}t}{1-\nu^2
  t^2}\frac{dt}{d\tau}\frac{dx^i}{d\tau}.
\end{equation}
In the spirit of Cartan, the two equations may be regarded as the
component ones of geodesic equation as long as what is called the
Newton-Cartan-like connection is taken:
\begin{eqnarray}
  & & \Gamma_{tt}^{\ t}=\frac{2\nu^{2}t}{1-\nu^{2}t^{2}}, \quad
  \Gamma_{tj}^{\ t}=0, \quad \Gamma_{ij}^{\ t}=0,
\label{t connection} \\
  & & \Gamma_{tt}^{\ i}=-\frac{F^i} m, \quad
  \Gamma_{tj}^{\ i}=\frac{\nu^{2}t}{1-\nu^{2}t^{2}}\delta_{j}^{i}, \quad
  \Gamma_{jk}^{\ i}=0.
\label{i connection}
\end{eqnarray}
Compared with the NH invariant connection in Appendix
\ref{app:NHinv}, only $\Gamma_{tt}^{\ i}$ among the
Newton-Cartan-like connection coefficients have different values.
Though one can easily see from Appendix \ref{app:NHinv} that the
other coefficients are still invariant under NH transformations in
spite of non-vanishing $\Gamma_{tt}^{\ i}$, one may suspect the
legality of the first equation in eq.(\ref{i connection}) because
$F^i$ transforms under NH transformations as acceleration does
(see eq.(\ref{a trans})) while $\Gamma_{tt}^{\ i}$ are connection
coefficients and have different transformation law in general.
Fortunately, they satisfy the same transformation law for NH
transformations, provided the other coefficients of the connection
are given as in eqs.(\ref{t connection},\ref{i connection}). The
transformation law of $\Gamma_{tt}^{\ i}$ under NH transformations
is discussed in detail in Appendix \ref{app:itt}.

If the above connection coefficients are chosen, the proper time
$\tau$ on empty $\mathcal{NH}$ is still acting as an affine
parameter in a gravitational field. It can also be verified that
in this case a geodesic tangent to the hypersurface of
simultaneity $t = t_0$ at $(t_0, \vect{x}_0)$ will not leave this
hypersurface.

For the gravitational field equation, we have three constraints:
the first is that its form must be invariant under NH
transformations; the second is that it must reduce to its
Newtonian counterpart when $\nu\to 0$; the third is that it must
reduce to the empty case (\ref{empty NC}) if there is no matter at
all. Thus we can assume the following form of this equation:
\begin{equation}\label{grav}
R_{tt}=4\pi G\varrho(\vect{x},t)g_{tt}-3\nu^{2}g_{tt},
\end{equation}
where $G$ is the Newton-like gravitational constant and the mass
density $\varrho(\vect{x},t)$ is a scalar under NH
transformations. The connection (\ref{t connection},\ref{i
connection}) gives the only non-vanishing components of curvature
tensor
\begin{equation}
R_{itt}^{\ \ j}\equiv -R_{tit}^{\ \ j}=\partial_{i}\Gamma_{tt}^{\
j}-\nu^{2}(1-\nu^{2}t^{2})^{-2}\delta_{i}^{j}
\end{equation}
and Ricci tensor
\begin{equation}
R_{tt}=\partial_{i}\Gamma_{tt}^{\
i}-3\nu^{2}(1-\nu^{2}t^{2})^{-2},
\end{equation}
where the second terms of both equations coincide with the empty
case (\ref{curv}). From eq.(\ref{grav}) we have the following
field equation:
\begin{equation}\label{eq:connection}
\partial_{i}\Gamma_{tt}^{\ i}=\frac{4\pi G\varrho(\vect{x},t)}{(1-\nu^2 t^2)^2}.
\end{equation}


\subsection{Law of Gravity for Spherical Source}

To solve eq.(\ref{eq:connection}), a curl-free condition must be
introduced as usual. This implies that $\Gamma_{tt}^{\ i}$ can be
expressed as the gradient of a scalar potential $V$ (cf Appendix
\ref{app:E}), which is responsible to the gravity induced by
compact objects:
\begin{equation}
\Gamma_{tt}^{\ i}(t,x)=\frac{\partial_i V(t,x)}{1-\nu^2 t^2}.
\end{equation}
Thus we have
\begin{equation}\label{Poisson}
\triangle V=4\pi G \varrho(\vect{x},t),
\end{equation}
where $\triangle:=\hat g^{ij} \r_i \r_j$ and $\hat g^{ij}$ is the
inverse of $\hat g_{ij}$ in eq.(\ref{l}). This equation has the
same form of Poisson equation for Newton's gravity.

For point-like gravitational source at $\vect{X}$, the mass
density has the form
\begin{equation}
\varrho(\vect{x},t)=(1-\nu^{2}t^{2})^{3
/2}M\delta^3(\vect{x}-\vect{X}),
\end{equation}
which is an NH scalar and comes back to the Newtonian case when
$\nu\to 0$. Here $M$ is the mass of the point-like source. (Such a
density is consistent with the density of probability from the NH
Schr\"odinger equation, as we can see in the next section.) For
boundary condition $V\to 0$ as $|\vect{x}|\to\infty$,
eq.(\ref{Poisson}) is straightforward to be solved with
\begin{equation}
V=-\frac{\sigma^{1/2}(t)G M}{|\vect{x}-\vect{X}|}.
\end{equation}
So the connection
\begin{equation}
\Gamma_{tt}^{\ i}=\frac{G
M}{\sigma^{1/2}(t)}\frac{x^i-X^i}{|\vect{x}-\vect{X}|^3}
\end{equation}
and the equation of motion for the test particle is obtained as
\begin{equation}\label{EOM}
\frac{d^{2}x^{i}}{dt^{2}}=-\frac{G
M}{\sigma^{1/2}(t)}\frac{x^{i}-X^{i}}{|\vect{x}-\vect{X}|^{3}}.
\end{equation}
Compared with the ordinary (Newton's) law of gravity, it is
interesting to see that the effect of Newton-Hooke parameter $\nu$
can be totally embodied by a time-dependent gravitational
``constant'' $\tilde{G}(t):=\sigma^{-1/2}(t)G$, at least from the
viewpoint of particle mechanics. One can check that the form of
eq.(\ref{EOM}) is invariant under NH transformations. We also
learn from this law of gravity that Newton-Hooke's third law holds
in this case.

In terms of the well-used coordinates on $\NH$ as homogeneous
space $N_+ (1,d)/\hat{N}_+ (1,d)$ \cite{DD,GP},\footnote{We call
them static coordinates for convenience.} whose relation to the
Beltrami coordinates is \cite{GHTXZ}:
\begin{eqnarray}
\tau&=&\nu\inv\tanh\inv\nu t, \label{tau} \\
q^i&=&\frac{x^i}{\sigma^{1/2}(t)}, \label{q}
\end{eqnarray}
eq.(\ref{EOM}) becomes
\begin{equation}
\frac{d^{2}q^{i}}{d\tau^{2}}-\nu^{2}q^{i}=-G
M\frac{q^{i}-Q^{i}}{|\vect{q}-\vect{Q}|^{3}}.
\end{equation}
This is exactly a particular case of the so-called
Dimitriev-Zel'dovich equation \cite{DZ,Peebles,GP}. As in the
usual (Newtonian) case, the result for point-like source can be
readily extended to the exterior of spherical source.


\section{Schr\"odinger Equation on the Newton-Hooke Space-time}
\label{sec:Sch}


\subsection{Schr\"odinger Equation from Geometric Contraction}

From the algebraic viewpoint, the usual Schr\"odinger equation can
be deduced from the second Casimir operator of the extended
Galilei algebra $\mathfrak{gal}^\mathrm{C}$. This standard method
can be applied to the NH case, which is shown in Appendix
\ref{app:SchEq}. Here we want to show how the Schr\"odinger
equation on $\NH$ can be directly obtained from a geometric
contraction. Rewrite the Klein-Gordon equation on dS spacetime in
terms of Beltrami coordinates \cite{GHTXZ} as
\begin{equation}
[\partial_{i}\partial_{i}-c^{-2}\partial_{t}\partial_{t}+R^{-2}(t^{2}\partial_{t}^{2}+2t
x^{i}\partial_{t}\partial_{i}+x^{i}x^{j}\partial_{i}\partial_{j}+2t\partial_{t}+2x^{i}\partial_{i})]\phi
=m^{2}c^{2}\sigma^{-1}\phi.
\end{equation}
In order to subtract the static energy, we substitute
\begin{equation}\label{phi}
\phi=\psi(\vect{x},t)e^{-\mathrm{i}m c^{2}f(t)}
\end{equation}
into the above equation and require the terms of order $c^2$ to
cancel out, which gives the condition
\begin{equation}
\frac{df}{dt}=(1-\nu^{2}t^{2})^{-1}.
\end{equation}
Noting eq.(\ref{dtau}) it is easy to see $f=\tau$ and the explicit
form is given by eq.(\ref{tau}), which makes eq.(\ref{phi}) of
clear physical meaning. Now omitting terms of order $c^{-2}$, we
obtain the following Schr\"odinger equation for free particle on
$\NH$:
\begin{equation}\label{free SchEq}
\mathrm{i}\partial_{t}\psi=\Big[{-\frac{\nabla^2}{2m}+\frac{\mathrm{i}\nu^{2}t
x^{i}\partial_{i}}{\sigma(t)}-\frac{m\nu^{2}
\vect{x}^{2}}{2\sigma^2(t)}}\Big]\psi.
\end{equation}

The invariance of eq.(\ref{free SchEq}) under NH transformations
is interesting, which actually gives the realization of the
extended NH group $N_+^\mathrm{C}$.\footnote{There is standard
method to obtain the realization of the extended group
\cite{DD,Bargmann}. The really interesting thing here is that the
local exponents are rational expressions in terms of the Beltrami
coordinates.} For simplicity, we consider rotation, time
translation, space translation and boost one by one. First,
eq.(\ref{free SchEq}) is obviously invariant under rotation if the
wave function $\psi$ is invariant. Second, time translation
(\ref{NH t}) gives an overall factor
$$\sigma(t_a)\sigma^{-2}(t_a,t)$$
to eq.(\ref{free SchEq}) for $\tilde{t}$ if the wave function
$\psi$ keeps invariant, so eq.(\ref{free SchEq}) is again
invariant. Third, the NH space translation
\begin{equation}\label{xtrans}
\tilde{x}^i=x^i-a^i
\end{equation}
needs some careful consideration. In fact, the wave function
cannot keep invariant in this case, in contrast to that of the
ordinary Schr\"odinger equation. It transforms as
\begin{equation}\label{psitrans}
\psi=\tilde{\psi}\exp[\mathrm{i}m\nu^{2}t(1-\nu^{2}t^{2})^{-1}(\vect{a}\cdot\tilde{\vect{x}}+\half\vect{a}^{2})].
\end{equation}
It is easy to check that its inverse transformation takes the same
form, which in fact imposes strong restriction on the possible
forms of the wave function transformation. The calculation to
substitute eqs.(\ref{xtrans},\ref{psitrans}) into eq.(\ref{free
SchEq}) and check the invariance is straightforward but a little
laborious. Finally, the case of boost
\begin{equation}
\tilde{x}^i=x^i-u^i t
\end{equation}
is even more complicated. Eq.(\ref{free SchEq}) turns out to be
invariant under this transformation when the wave function
transforms as
\begin{equation}\label{psiboost}
\psi=\tilde{\psi}\exp[\mathrm{i}m(1-\nu^{2}t^{2})^{-1}(\vect{u}\cdot\tilde{\vect{x}}+\half\vect{u}^{2}t)].
\end{equation}
As expected, the wave function transformations (\ref{psitrans})
and (\ref{psiboost}) come back to their familiar forms when
$\nu=0$, which gives the Galilean invariance of the ordinary
Schr\"odinger equation. Because an arbitrary NH transformation can
be composed by the above transformations, we have verified the
Newton-Hooke invariance of eq.(\ref{free SchEq}).

To introduce interactions into Schr\"odinger equation (\ref{free
SchEq}), one just add a term to it:
\begin{equation}\label{eq:SchEq}
\mathrm{i}\partial_{t}\psi=\Big[{-\frac{\nabla^2}{2m}+\frac{\mathrm{i}\nu^{2}t
x^{i}\partial_{i}}{\sigma(t)}-\frac{m\nu^{2}
\vect{x}^{2}}{2\sigma^2(t)}+\frac{U(\vect{x},t)}{\sigma(t)}}\Big]\psi,
\end{equation}
where $U(\vect{x},t)$ is a real scalar under NH transformations.
Then one can easily check the NH invariance of this Schr\"odinger
equation based on the above discussion.


\subsection{Conservation of Probability}

It is interesting to ask whether there is something for
eq.(\ref{eq:SchEq}) corresponding to the conservation of
probability for the ordinary Schr\"odinger equation. Take the
complex conjugation of eq.(\ref{eq:SchEq}):
\begin{equation}\label{SchCC}
-{\rm i}\partial_{t}\psi^*=\Big[{-\frac{\nabla^2}{2m}-\frac{{\rm
i}\nu^{2}t
x^{i}\partial_{i}}{\sigma(t)}-\frac{m\nu^{2}\vect{x}^2}{2\sigma^2(t)}+\frac{U(\vect{x},t)}{\sigma(t)}}\Big]\psi^*.
\end{equation}
By constructing
$\sigma^{-3/2}(t)[\psi^*\times$(\ref{eq:SchEq})$-\psi\times$(\ref{SchCC})]
and rearranging it, we get
\begin{equation}
\partial_{t}\Big[{\sigma^{-3/2}(t)\,\psi^*\psi}\Big] = \nabla\cdot\Big[{\sigma^{-3/2}(t)\frac{{\rm
i}}{2m}(\psi^*\nabla\psi -\psi\nabla\psi^*)
+\sigma^{-5/2}(t)\nu^{2}t(\psi^*\vect{x}\psi)}\Big]. \nno
\end{equation}
So if we define the density of probability as
\begin{equation}\label{rho}
\rho=\sigma^{-3/2}(t)\psi^*\psi,
\end{equation}
and the flux of probability as
\begin{equation}\label{j}
\vect{j}=\sigma^{-3/2}(t)\frac{{\rm
i}}{2m}(\psi\nabla\psi^*-\psi^*\nabla\psi)
-\sigma^{-5/2}(t)\nu^{2}t(\psi^*\vect{x}\psi),
\end{equation}
we do have something like the conservation of probability:
\begin{equation}\label{prob}
\partial_{t}\rho+ \nabla\cdot\vect{j}=0.
\end{equation}

In fact, one can check that the expression $\frac{{\rm
i}}{2m}(\psi\nabla\psi^*-\psi^*\nabla\psi)$ in eq.(\ref{j}) has
the same NH transformation property (\ref{j trans}) as
$\vect{\jmath}$ defined in the NH continuous medium mechanics. So
it is easy to see from the NH invariance of $\psi^*\psi$ that
$\rho$ and $\vect{j}$ defined here have the same NH transformation
properties as $\hat{\rho}$ and $\hat{\vect{j}}$ in the NH
continuous medium mechanics, respectively, and that
eq.(\ref{prob}) can be regarded as the quantum correspondence of
the equation of continuity (\ref{flat cont}). This correspondence
justifies eq.(\ref{prob}) as the genuine equation for the
conservation of probability.


\section{Conclusion and Discussion}
\label{sec:CD}

In this article we have mainly discussed the dynamical aspects on
Newton-Hooke space-time and established the consistent theory of
gravity on this space-time as a kind of Newton-Cartan-like theory.
In our discussion, we concentrate on the geometric properties of
NH space-time and the NH invariance of physics on it. We obtain
the NH space-time manifold, NH transformation (\ref{NH t},\ref{NH
x}) and NH Schr\"odinger equation (\ref{eq:SchEq}) {\it etc}
directly from In\"on\"u-Wigner contraction of their dS
counterparts under the Beltrami coordinates. For the NH quantum
mechanics, we find that a conservative probability can be defined
and related to the mass density in NH fluid mechanics.

For NH space-time the two most useful coordinates are the Beltrami
coordinates and the static coordinates, their relation being
eqs.(\ref{tau},\ref{q}). It is interesting to see from the NH
transformation (\ref{NH t},\ref{NH x}) that in the Beltrami
coordinates NH space-time is spatially uniform, while in the
static coordinates it is temporally uniform. The Beltrami
coordinates are introduced through some projective-geometry-like
method \cite{GHXZ, GHXZ2, GHTXZ5}. It is not strange that
Beltrami-dS spacetime or NH space-time has something to do with
projective geometry, since there is systematic
projective-geometry-like method to deal with constant curvature
spaces \cite{GHTXZ4}. If the so-called ``elliptic" interpretation
of dS spacetime \cite{elliptic}, {\it i.e.}, dS spacetime with
topology $\hypb/\mathbb{Z}_2$, is taken, one can examine dS
spacetime really from projective geometry point of view. It should
be mentioned that the key difference between $\hypb$ and
$\hypb/\mathbb{Z}_2$ is that the latter is not orientable while
the former is orientable.

For the Newton-Cartan-like theory on NH space-time, unlike the
previous papers that take serious the diffeomorphism invariance,
we concentrate on the NH invariance and construct a theory of
gravity which preserves the NH invariance of the formulation of
physical laws and of as many as possible coefficients of the
connection. We see from eqs.(\ref{t connection},\ref{i
connection}) that in the Beltrami coordinates the contributions
from the cosmological background and material gravitation to the
connection are completely decoupled, while in other coordinates,
say the static coordinates (\ref{tau},\ref{q}), they are not. One
can reasonably expected that Beltrami coordinate systems are the
only one having this property, for there exists Newton-Hooke's
first law, {\it i.e.}, the law of inertia.

The discussion in this article can be easily applied to
Beltrami-AdS spacetime and the corresponding $\NH_-$. It is also
readily extended to space-time dimensions other than four.
Especially, our Newton-Cartan-like formalism in Sec.\ref{sec:NC}
can be contracted to the Newton-Cartan theory on the Galilei
space-time as the case of $\nu \to 0$.

\begin{acknowledgments}
The authors would like to thank Professors Q.-K.~Lu, J.-Z.~Pan and
X.-C.~Song for valuable discussions. Y. Tian would also like to
thank Dr. H.-Z.~Chen for helpful suggestions. This work is partly
supported by NSFC under Grant Nos. 90103004, 10175070, 10375087,
10347148, 10373003 and 90403023.
\end{acknowledgments}

\appendix


\section{Newton-Hooke Invariant Connection \label{app:NHinv}}

In this appendix, we investigate connections that are invariant
under NH transformations. We shall prove that the connection
(\ref{affine}) contracted from Beltrami-dS spacetime is the only
NH invariant connection that keeps the proper time $d\tau$
invariant under NH transformations.

By the term NH invariant connection, we refer to a
connection with coefficients depending on the
Beltrami coordinates in the same way under NH
transformations:
\begin{equation}
  \tilde{\Gamma}_{\mu\nu}^{\ \rho}(\tilde{x})
  = \Gamma_{\mu\nu}^{\ \rho}(\tilde{x}),
\label{eq:GammaInv}
\end{equation}
while, at the same time, the transformation law
\begin{equation}
  \tilde{\Gamma}_{\mu\nu}^{\ \rho}(\tilde{x})
= \frac{\partial\tilde{x}^\rho}{\partial x^\sigma}\,
  \frac{\partial x^\alpha}{\partial\tilde{x}^\mu}\,
  \frac{\partial x^\beta}{\partial\tilde{x}^\nu}\,
  \Gamma_{\alpha\beta}^{\ \sigma}(x)
+ \frac{\partial\tilde{x}^\rho}{\partial x^\sigma}\,
  \frac{\partial^2 x^\sigma}{\partial\tilde{x}^\mu\partial\tilde{x}^\nu}
\label{eq:GammaTransf}
\end{equation}
is satisfied.

First, space translation \{$\tilde{t} = t$, $\tilde{x}^i =
x^i - b^i$\} results in $\frac{\partial\tilde{x}^\mu}{\partial x^\nu}
= \delta^\mu_\nu$ and thus
$ \tilde{\Gamma}_{\mu\nu}^{\ \rho}(\tilde{x}) = \Gamma_{\mu\nu}^{\ \rho}(x)$
according to eq.(\ref{eq:GammaTransf}).
Substituting it into eq.(\ref{eq:GammaInv}), we immediately obtain
$\Gamma_{\mu\nu}^{\ \rho}(t, \vect{x} - \vect{b})
 = \Gamma_{\mu\nu}^{\ \rho}(t, \vect{x}),$
which implies that each $\Gamma_{\mu\nu}^{\ \rho}$ depends on $t$ only.
Next, for boosts \{$\tilde{t} = t$, $\tilde{x}^i = x^i - u^i t$\},
\[
  \frac{\partial\tilde{t}}{\partial t} = 1, \quad
  \frac{\partial\tilde{t}}{\partial x^i} = 0, \quad
  \frac{\partial\tilde{x}^i}{\partial t} = - u^i, \quad
  \frac{\partial\tilde{x}^i}{\partial x^j} = \delta^i_j,
\]
and inversely,
\[
  \frac{\partial t}{\partial\tilde{t}} = 1, \quad
  \frac{\partial t}{\partial\tilde{x}^i} = 0, \quad
  \frac{\partial x^i}{\partial\tilde{t}} = u^i, \quad
  \frac{\partial x^i}{\partial\tilde{x}^j} = \delta^i_j.
\]
The transformation law (\ref{eq:GammaTransf})
gives rise to
\be 
\tilde{\Gamma}_{tt}^{\ t}(\tilde{t})
  &=& \Gamma_{tt}^{\ t}(t) + 2\,\Gamma_{tj}^{\ t}(t)\,u^j
  + \Gamma_{ij}^{\ t}(t)\,u^i u^j , \nno \\
  \Gamma_{tt}^{\ i}(t) &=& -\Gamma_{tt}^{\ t}(t)\, u^i
  + \Gamma_{tt}^{\ i}(t) + 2\,\Gamma_{jt}^{\ i}(t)\,u^j
  + \Gamma_{jk}^{\ i}(t)\,u^j u^k. \nno
\ee 
On the other hand, the invariance of the connection indicates
$\tilde{\Gamma}_{tt}^{\ t}(\tilde{t}) = \Gamma_{tt}^{\
t}(\tilde{t}) = \Gamma_{tt}^{\ t}(t)$ and $\tilde{\Gamma}_{tt}^{\
i}(\tilde{t}) = \Gamma_{tt}^{\ i}(\tilde{t}) = \Gamma_{tt}^{\
i}(t)$. These, together with the above results, imply that
\be 
\label{eq:ttj-tij}
  \Gamma_{tj}^{\ t}(t) &=& \Gamma_{jt}^{\ t}(t) = 0, \qquad
  \Gamma_{ij}^{\ t}(t) = 0, \\
  \Gamma_{jk}^{\ i}(t) &=& 0, \qquad
  \Gamma_{tj}^{\ i}(t) =
  \Gamma_{jt}^{\ i}(t) = \frac{1}{2}\,\Gamma_{tt}^{\ t}(t)\,\delta^i_j.
\ee 
As for $\Gamma_{tt}^{\ t}(t)$, its transformation law and
invariance under NH time translation indicate
\begin{equation}\label{eq:ttt}
  \Gamma_{tt}^{\ t}\left(\frac{t - t_a}{1 - \nu^2 t_a\,t}\right)
  = \frac{(1 - \nu^2 t_a\,t)^2}{1 - \nu^2 t_a^2}\,\Gamma_{tt}^{\ t}(t)
  - \frac{2\nu^2 t_a\,(1 - \nu^2 t_a\,t)}{1 - \nu^2 t_a^2}.
\end{equation}
Taking the derivative of the above equation with respect to $t_a$ at $t_a = 0$, then we have 
\begin{equation}
  (1 - \nu^2 t^2)\,\frac{d\Gamma_{tt}^{\ t}}{dt}
  = 2\nu^2 + 2\nu^2 t\,\Gamma_{tt}^{\ t}.
\end{equation}
The general solution of this ODE reads
\begin{equation}\label{eq:inv ttt}
  \Gamma_{tt}^{\ t}(t) = \frac{2\nu^2 t + 2C\nu}{1 - \nu^2 t^2}
\end{equation}
with $C$ the integral constant. Finally, under space rotations
\{$\tilde{t} = t$, $\tilde{x}^i = O^i_{\ j}\,x^j$\}, the
transformation law (\ref{eq:GammaTransf}) reduces to
$\tilde{\Gamma}_{tt}^{\ i}(\tilde{t}) = O^i_{\ j}\,\Gamma_{tt}^{\
j}(t)$. Due to the invariance, we have $\Gamma_{tt}^{\ i}(t) =
O^i_{\ j}\,\Gamma_{tt}^{\ j}(t)$ for arbitrary $(O^i_{\ j})\in
SO(3,\mathbb{R})$. This is only possible when
\begin{equation}
  \Gamma_{tt}^{\ i} = 0.
\end{equation}

The above NH invariant connection is almost the same as that
(\ref{affine}) contracted from Beltrami-dS spacetime, except for
an arbitrary integral constant $C$. It is easy to prove that the
NH invariance of proper-time element $d\tau = \si^{-1}(t)dt$
requires $C=0$, because the first integral of the temporal
component of the geodesic equation
\begin{equation}\label{t geodesic}
\frac{d^{2}t}{d\tau^{2}}+\Gamma_{\mu\nu}^{\
t}(t,x)\frac{dx^{\mu}}{d\tau}\frac{dx^{\nu}}{d\tau}=0
\end{equation}
is
\begin{equation}\label{dt}
\frac{dt}{d\tau}=(1-\nu^{2}t^{2})\left(\frac{1 -\nu t}{1 +\nu
t}\right)^C,
\end{equation}
which leads to $C=0$.


\section{Energy in Newton-Hooke Mechanics}
\label{app:E}

In our formulism, the manifest time-translation invariance is
lost. However, since there is ``timelike" Killing vector
$\mathbf{H}$ (\ref{realize}) in $\NH$ (which is $\partial_\tau$ in
terms of the coordinates (\ref{tau},\ref{q}), actually), we can
expect that some kind of energy conservation law should exist.
Before investigating the energy conservation law, we first give
some justifications for the kinetic energy (\ref{energy}) obtained
from contraction of Beltrami-dS spacetime in \cite{GHTXZ}. Under
coordinate transformation (\ref{tau},\ref{q}) it becomes
\begin{equation}
E_\mathrm{k}=\frac{m}{2}\Big(\frac{d\vect{q}}{d\tau}\Big)^{2}-\half
m\nu^{2}\vect{q}^{2},
\end{equation}
which is just the (conservative) total energy of an anti-harmonic
oscillator, as is well known as the conservative energy in (empty)
$\NH$ \cite{DD,Gao}. It has obvious $\tau$-translational
invariance. For its full NH transformation property, we can
consider $\vect{v}^{2}-\nu^{2}(\vect{x}-\vect{v}t)^{2}$ from
eq.(\ref{energy}). A lengthy but straightforward calculation gives
\begin{equation}
\tilde{\vect{v}}^{2}-\nu^{2}(\tilde{\vect{x}}-\tilde{\vect{v}}\tilde{t})^{2}
=(\vect{v}-\vect{u})^{2}-\nu^{2}(\vect{x}-\vect{b}-\vect{v}t)^{2},
\end{equation}
which is elegant and whose $t_a$-independence is what we wanted.

Let us write down the total energy $E$ as
\begin{equation}
E=E_\mathrm{k}+V,
\end{equation}
where $V$ is the potential energy. We require the conservation of
$E$ along the world line, which gives from eq.(\ref{kinetic})
\begin{equation}\label{dE/ds}
\frac{dE}{dt}=[(1-\nu^{2}t^{2})F^i+\partial_i
V]\frac{dx^i}{dt}+(\nu^{2}t x^i F^i+\partial_t V)=0.
\end{equation}
Since this equation is valid for arbitrary $d\vect{x}/dt$, we will
have
\begin{eqnarray}
(1-\nu^{2}t^{2})F^i+\partial_i V&=&0, \label{potential} \\
\nu^{2}t x^i F^i+\partial_t V&=&0,
\end{eqnarray}
if the velocity-independence of both
$\vect{F}=\vect{F}(t,\vect{x})$ and $V=V(t,\vect{x})$ is assumed.
Eq.(\ref{potential}) is a generalization of the usual relation
$\partial_i V=-F^i$.

Thus, we have the equation
\begin{equation}\label{VPDE}
  \frac{\partial V}{\partial t} - \frac{\nu^2 t}{1-\nu^2 t^2}x^i\partial_i V
  = 0
\end{equation}
for $V$. This PDE can be solved with general solution
\begin{equation}
V=V(\sigma^{-1/2}(t)\vect{x}).
\end{equation}
Noticing eq.(\ref{q}), it is
\begin{equation}
V=V(\vect{q}),
\end{equation}
which is reasonable because of the manifest time-translation
invariance in coordinates (\ref{tau},\ref{q}).


\section{Transformation Property of $\Gamma_{tt}^{\ k}$}
\label{app:itt}

It is necessary and interesting to investigate
$\Gamma_{tt}^{\ k}$ in eq.(\ref{i connection}).
The $k$-$tt$ component equation of transformation law
Eq.(\ref{eq:GammaTransf}), the first equation of (\ref{t
connection}), and the latter two equations of (\ref{i connection})
give
\[
\tilde{\Gamma}_{tt}^{\ k}(\tilde{t},\tilde{x})=\Gamma_{tt}^{\
i}(t,x)\frac{\partial t}{\partial\tilde{t}}\frac{\partial
t}{\partial\tilde{t}}\frac{\partial\tilde{x}^{k}}{\partial
x^{i}}+\frac{2\nu^{2}t}{1-\nu^{2}t^{2}}\frac{\partial
t}{\partial\tilde{t}}\Big(\frac{\partial
t}{\partial\tilde{t}}\frac{\partial\tilde{x}^{k}}{\partial
t}+\frac{\partial
x^{i}}{\partial\tilde{t}}\frac{\partial\tilde{x}^{k}}{\partial
x^{i}}\Big)+\frac{\partial^{2}x^{\rho}}{\partial\tilde{t}\partial\tilde{t}}
\frac{\partial\tilde{x}^{k}}{\partial x^{\rho}}.
\]
The second term on the RHS vanishes since
\begin{equation}
\frac{\partial
t}{\partial\tilde{t}}\frac{\partial\tilde{x}^{k}}{\partial
t}+\frac{\partial
x^{i}}{\partial\tilde{t}}\frac{\partial\tilde{x}^{k}}{\partial
x^{i}}=\frac{\partial\tilde{x}^{k}}{\partial\tilde{t}}=0.
\end{equation}
The third term on the RHS actually includes two parts:
\begin{equation}
\frac{\partial^{2}t}{\partial\tilde{t}\partial\tilde{t}}
\frac{\partial\tilde{x}^{k}}{\partial
t}+\frac{\partial^{2}x^i}{\partial\tilde{t}\partial\tilde{t}}
\frac{\partial\tilde{x}^{k}}{\partial x^i},
\end{equation}
which exactly cancels each other, as shown by careful calculation.
The following identity is useful to this calculation:
\begin{equation}
\sigma(t_a,t)\sigma(-t_a,\tilde{t})=\sigma(t_a).
\end{equation}

Thus, the NH transformation of $\Gamma_{tt}^{\ k}$ has the
following form:
\begin{equation}
\tilde{\Gamma}_{tt}^{\ k}(\tilde{t},\tilde{x})=\Gamma_{tt}^{\
i}(t,x)\frac{\partial t}{\partial\tilde{t}}\frac{\partial
t}{\partial\tilde{t}}\frac{\partial\tilde{x}^{k}}{\partial
x^{i}}= \frac{\sigma^3(t_a,t)}{\sigma^{3/2}(t_a)}O^k_{\
i}\Gamma_{tt}^{\ i},
\end{equation}
which is, in fact, the same as that of acceleration (\ref{a
trans}). This justifies the first equation in eq.(\ref{i
connection}).


\section{Schr\"odinger Equation from Algebraic Viewpoint}
\label{app:SchEq}

As is well known, the familiar Schr\"odinger equation can be
written as
\begin{equation}\label{Schrodinger}
C_2\psi(\vect{x},t)=2m U(\vect{x},t)\psi(\vect{x},t),
\end{equation}
where $C_2$ is the second order Casimir operator of the central
extension $\mathfrak{gal}^\mathrm{C}$ of Galilei algebra:
\begin{equation}
C_2=2\mathrm{i}m\mathbf{H}+\mathbf{P}^2,
\end{equation}
and $U(\vect{x},t)$ is a scalar under NH transformations. Since
eq.(\ref{Schrodinger}) satisfies the requirement of symmetry and
is rather general, the Schr\"odinger equation in $\mathcal{NH}$
should also be given by it. For the extended NH algebra
$\mathfrak{n}_+^\mathrm{C}$, the second Casimir is
\begin{equation}
C_2=2\mathrm{i}m\mathbf{H}+\mathbf{P}^2-\nu^2\mathbf{K}^2,
\end{equation}
where the realization (\ref{realize}) is modified to
\begin{equation}
\mathbf{P}_i=\partial_i-\frac{\mathrm{i}m\nu^2 t x^i}{1 - \nu^2
t^2}, \quad \mathbf{K}_i=t\partial_i-\frac{\mathrm{i}m x^i}{1 -
\nu^2 t^2}.
\end{equation}
It is straightforward to check that the above realization
satisfies the $\mathfrak{n}_+^\mathrm{C}$ algebra (\ref{NH algebra
ext}). After a little calculation, one obtains the Schr\"odinger
equation on $\mathcal{NH}$ same as eq.(\ref{eq:SchEq}).

\end{document}